\documentclass[a4paper,twocolumn,showpacs]{revtex4}

\usepackage[english]{babel}
\usepackage{graphicx}
\usepackage{xspace}
\usepackage{longtable}
\usepackage{amsmath,amssymb}

\DeclareMathAlphabet{\mathbfit}{OT1}{cmr}{bx}{it}

\begin{document}

\title{X-ray magnetic circular dichroism study of Re ${\mathbfit{5d}}$ magnetism in Sr$_{\mathbf 2}$CrReO$_{\mathbf 6}$}

\author{P.~Majewski}
\email{Petra.Majewski@wmi.badw.de}
\affiliation{Walther-Mei{\ss}ner-Institut, Bayerische Akademie der
Wissenschaften, Walther-Mei{\ss}ner Str.~8, 85748 Garching, Germany}
\author{S.~Gepr\"{a}gs}
\affiliation{Walther-Mei{\ss}ner-Institut, Bayerische Akademie der
Wissenschaften, Walther-Mei{\ss}ner Str.~8, 85748 Garching,
Germany}
\author{O.~Sanganas}
\affiliation{Walther-Mei{\ss}ner-Institut, Bayerische Akademie der
Wissenschaften, Walther-Mei{\ss}ner Str.~8, 85748 Garching, Germany}
\author{M.~Opel}
\affiliation{Walther-Mei{\ss}ner-Institut, Bayerische Akademie der
Wissenschaften, Walther-Mei{\ss}ner Str.~8, 85748 Garching, Germany}
\author{R.~Gross}
\affiliation{Walther-Mei{\ss}ner-Institut, Bayerische Akademie der
Wissenschaften, Walther-Mei{\ss}ner Str.~8, 85748 Garching, Germany}
\author{F.~Wilhelm}
\affiliation{European Synchrotron Radiation Facility (ESRF), 6 Rue
Jules Horowitz, BP 220, 38043 Grenoble Cedex 9, France}
\author{A.~Rogalev}
\affiliation{European Synchrotron Radiation Facility (ESRF), 6 Rue
Jules Horowitz, BP 220, 38043 Grenoble Cedex 9, France}
\author{L.~Alff}
\email{alff@oxide.tu-darmstadt.de} \affiliation{Darmstadt
University of Technology, Petersenstr.~23, 64287 Darmstadt,
Germany}

\date{\today}

\begin{abstract}

We have measured Re $5d$ spin and orbital magnetic moments in the
ferrimagnetic double perovskite Sr$_2$CrReO$_6$ by X-ray magnetic
circular dichroism at the $L_{2,3}$ edges. In fair agreement with
recent band-structure calculations \cite{Vaitheeswaran:05}, at the
Re site a large $5d$ spin magnetic moment of
$-0.68\,\mu_\textrm{B}$ and a considerable orbital moment of
$+0.25\,\mu_\textrm{B}$ have been detected. We found that the
Curie temperature of the double perovskites $A_2BB'$O$_6$ scales
with the spin magnetic moment of the 'non-magnetic' $B'$ ion.

\end{abstract}
\maketitle

Among the ferrimagnetic double perovskites, which are currently
considered as possible spintronics materials, the compound
Sr$_2$CrReO$_6$ with its Curie temperature $T_\textrm{C}$ of about
635\,K has the highest transition temperature observed so far in
this class of materials \cite{Kato:02,Asano:04,Kato:04}. The
physical mechanism leading to the high Curie temperature is still
unclear. Compared to the related double perovskites Sr$_2$CrWO$_6$
\cite{Philipp:01,Philipp:03} and Sr$_2$FeReO$_6$
\cite{Kobayashi:99,Auth:04} with $T_\textrm{C}\approx460$\,K and
$T_\textrm{C}\approx400$\,K respectively, Sr$_2$CrReO$_6$ has a
considerably increased $T_\textrm{C}$. From a recent neutron
diffraction study of the mixed compounds
Sr$_2$Fe$_{1-x}$Cr$_x$ReO$_6$, it was speculated that the cation
size matching improves with increasing $x$. This in turn increases
the $pd$ hybridization, contributing to an enhancement of
$T_\textrm{C}$ \cite{DeTeresa:05}. These observations can at least
qualitatively be understood within a generalized double exchange
or kinetic energy driven exchange model proposed by Sarma {\em et
al}.~\cite{Sarma:00}, where the itinerant electrons mediate an
indirect antiferromagnetic coupling between the Cr or Fe and the W
or Re moments.

It is still an open question whether the magnetic moments at the
'non-magnetic' W/Re sites are {\em induced} by the hybridization
with the magnetic 3d ion or {\em intrinsic} $5d$ moments. It has
been suggested recently that the critical temperature in this type
of ferrimagnetic double perovskites scales with the magnetic
moment of the 'non-magnetic' site \cite{Majewski:pre,Sikora:pre}.
An important experimental test for this idea is the investigation
of the spin and orbital magnetic moments at the Re site in
Sr$_2$CrReO$_6$ with its unexpectedly high transition temperature.
Band-structure calculations based on the full-potential linear
muffin-tin orbital method predict a value for the Re spin magnetic
moment of $-0.85\,\mu_\textrm{B}$ or $-0.69\,\mu_\textrm{B}$ using
the generalized gradient approximation (GGA) or local spin density
approximation (LSDA), respectively, both with included spin-orbit
(SO) coupling \cite{Vaitheeswaran:05}. First experimental results
from neutron scattering indicate a much smaller value of about
$-0.21\,\mu_\textrm{B}$ at 5\,K \cite{DeTeresa:05}. However, it is
of great importance to measure directly magnetic moments with a
state-of-the-art method as X-ray magnetic circular dichroism
(XMCD). The great advantage of XMCD is its element selectivity and
the possibility to extract spin and orbital magnetic moments using
the magneto-optical sum-rules \cite{Thole:92,Carra:93}. A
considerable orbital magnetic moment on Re is also predicted by
theory ($0.18\,\mu_\textrm{B}$), indicating the relevance of
spin-orbit-coupling which leads, in turn, to a break-down of the
complete half-metallicity in Sr$_2$CrReO$_6$
\cite{Vaitheeswaran:05}. In contrast, in Sr$_2$CrWO$_6$
\cite{Philipp:03,Majewski:pre} and Sr$_2$FeMoO$_6$
\cite{Kobayashi:98} the half-metallic character is preserved.

In this letter, we present a XMCD study of the Re moments in
Sr$_2$CrReO$_6$ polycrystalline bulk samples supplemented by SQUID
(superconducting quantum interference device) magnetization
measurements. Our measurements give further evidence that the
Curie temperature $T_\textrm{C}$ in the ferrimagnetic double
perovskites of the form $A_2BB'$O$_6$ with $A$ an alkaline earth,
$B$ a magnetic metal, and $B'$ an originally non-magnetic $3d$ or
$5d$ metal is strongly dependent on the magnetic moment at the
$B'$ site.

The samples were made from a stoichiometric mixture of SrO$_2$,
Re, and CrO$_2$ powders encapsulated in an evacuated silica tube.
The powder was sintered with temperature increased stepwise up to
850$^\circ$C before pressed to a pellet. This pellet was again
sintered in an evacuated silica tube at 1150$^\circ$C for four
days using Ni as getter material. The final samples contain only a
tiny amount of elemental Re as seen by x-ray diffraction using a
four circle geometry. From Rietveld-analysis we estimate the
amount of elemental Re less than 3\%. The amount of Cr/Re
antisites is about 9\%. From SQUID measurements we obtained
$T_\textrm{C}=635$\,K for Sr$_2$CrReO$_6$ and a saturation
magnetization $M_{\text{sat}}\approx0.89\,\mu_{\text{B}}$/f.u..

The XMCD measurements on the Re $L_{2,3}$ edges were performed at
the European Synchrotron Radiation Facility (ESRF) at beam line
ID12 \cite{Rogalev:01}. The spectra were recorded using the total
fluorescence yield detection mode. The XMCD spectra were obtained
as direct difference between consecutive XANES scans (X-ray
Absorption Near Edge Spectrum) recorded with opposite helicities
of the incoming X-ray beam. To ensure that the XMCD spectra are
free from experimental artefacts the data was collected for both
directions of the applied magnetic field of 6\,T (parallel and
antiparallel to the X-ray beam). The measurements were performed
at about 10\,K. Since the samples measured in backscattering
geometry were very thick, the spectra were corrected for
self-absorption effects. The edge jump ratio $L_3/L_2$ was
normalized to $2.20/1$ \cite{Henke}. This takes into account the
difference in the radial matrix elements of the $2p_{1/2}$ to
$5d(L_2)$ and $2p_{3/2}$ to $5d(L_3)$ transitions.

\begin{figure}
\centering{%
\includegraphics[width=0.8\columnwidth,clip=]{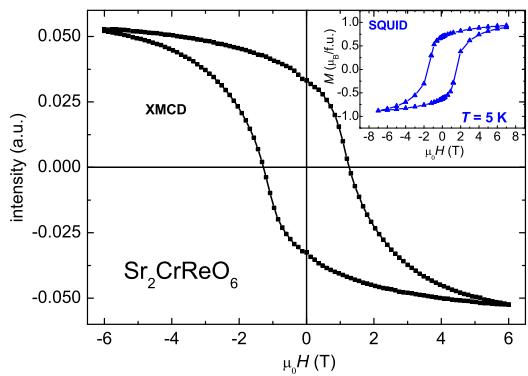}}
 \caption{(Color online) XMCD signal of Sr$_2$CrReO$_6$ versus applied magnetic field measured at the Re $L_2$
 edge at $T\approx10$\,K. The inset shows the
 magnetization vs.~magnetic field for this sample measured by SQUID at $T=5$\,K.}
 \label{Fig:hysterese}
\end{figure}

In Fig.~\ref{Fig:hysterese} we show the Re $5d$ magnetization
curve in Sr$_2$CrReO$_6$ at $T \approx 10$\,K measured by XMCD at
the Re $L_2$ absorption edge. A remarkably large coercive field of
about 1.27\,T is observed. The XMCD signal is in good agreement
with the macroscopic magnetization measured by SQUID at 5\,K (see
inset of Fig.~\ref{Fig:hysterese}) and other SQUID measurements
\cite{Kato:04}. The saturation magnetization of
0.89\,$\mu_\textrm{B}$ per formula unit is quite close to the
value of 1\,$\mu_\textrm{B}$ expected from a simple ionic picture,
where the magnetic moments of Cr$^{3+}$ and Re$^{5+}$ are coupled
antiferromagnetically.

\begin{figure}[tb]
\centering{\includegraphics[width=0.9\columnwidth,clip=]{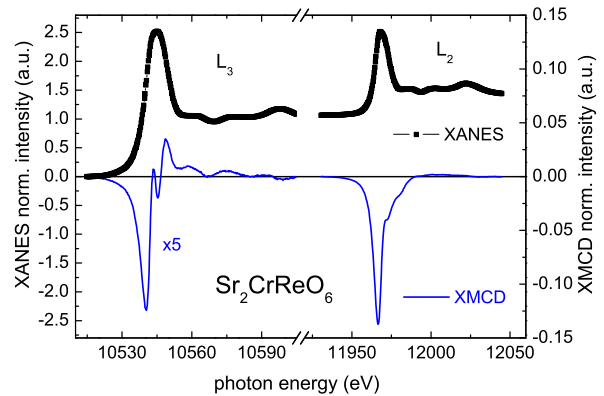}}
 \caption{(Color online) Re $L_{2/3}$ XANES (symbols) and XMCD (full line) spectra for Sr$_2$CrReO$_6$ recorded at $T\approx10$\,K and $B=6\,$T.
 The normalized XANES spectra are corrected for self-absorption effects.
 The XMCD L$_3$ spectrum is multiplied by a factor of 5 for better visibility.}
 \label{Fig:XMCD}
\end{figure}

We now discuss the XANES spectra of Sr$_2$CrReO$_6$ shown in
Fig.~\ref{Fig:XMCD}. The white lines at the Re $L_{2,3}$ edges
have a fine structure which reflects the Re $5d$ density of
unoccupied states influenced by the crystal field. The white lines
at both edges show a faint shoulder on the high energy side. This
can be interpreted as the signature of the crystal field splitting
of the $5d$ band into $t_\textrm{2g}$ and $e_\textrm{g}$ states
($\sim3$\,eV). More pronounced double peak structures of similar
splitting have been already observed at the Mo $L_{2,3}$ and W
$L_{2,3}$ edges for the double perovskites Sr$_2$FeMoO$_6$
\cite{Besse:02} and $A_2$CrWO$_6$ ($A=$\,Ca,~Sr)
\cite{Majewski:pre}.

\begin{table}[b]
\caption{\label{tab:table1} Summary of magnetic moments at the
'non-magnetic' ion Re in Sr$_2$CrReO$_6$ ($T_\textrm{C}= 635$\,K)
measured by XMCD using $n_{\text{h}}=5.3$. For comparison
calculated values with different approximations to the exchange
integral are also shown.}
\begin{ruledtabular}
\begin{tabular}{lccc}
 & $m_{\textrm{S}}$ ($\mu_{\textrm{B}}$/Re) & $m_{\textrm{L}}$ ($\mu_{\textrm{B}}$/Re) &
 $|m_{\textrm{L}}/m_{\textrm{S}}|$
 \\ \hline
experiment & -0.68 & 0.25 & 0.37\\
LSDA+SO \cite{Vaitheeswaran:05} & -0.69 & 0.17 & 0.25\\
GGA+SO \cite{Vaitheeswaran:05} & -0.85 & 0.18 & 0.21\\
\end{tabular}
\end{ruledtabular}
\end{table}

As shown in Fig.~\ref{Fig:XMCD}, for both absorption edges we find
a rather intense XMCD signal. This is a clear evidence for the
existence of a magnetic moment at the Re $5d$ shell. We have
extracted spin and orbital moments, $m_{\text{S}}$ and
$m_{\text{L}}$, of Re and summarized in Table~\ref{tab:table1}.
Note that in order to obtain the spin and orbital magnetic moments
separately the data must be normalized to the number of $5d$ holes
$n_{\text{h}}$. According to band-structure calculations we use
$n_{\text{h}}=5.3$ \cite{Delin:pc}. This has to be kept in mind
when comparing data, as $n_{\text{h}}$ is proportional to the
obtained moments. Only the ratio $m_{\text{L}}/m_{\text{S}}$ is
independent of the estimate of $n_{\text{h}}$. Both,
$m_{\text{L}}/m_{\text{S}}$ and $m_{\text{L}}$, are clearly larger
than expected from the calculations including spin-orbit coupling.
This underlines the importance of relativistic effects for the
heavy Re.

\begin{figure}[tb]
\centering{\includegraphics[width=0.8\columnwidth,clip=]{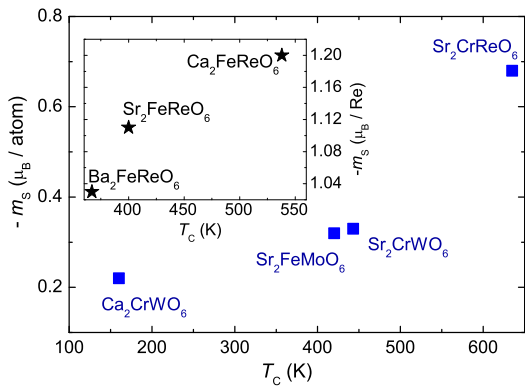}}
 \caption{(Color online) $B'$ spin magnetic moments for different double
 perovskites $A_2BB'$O$_6$ measured by XMCD on Mo \cite{Besse:02},
 our data on W \cite{Majewski:pre} and Re data presented in this paper. The number of $5d$ holes has been obtained from
 recent band-structure calculations \cite{Vaitheeswaran:05,Majewski:pre,Delin:pc}. In the inset
 further data deduced from XMCD \cite{Sikora:pre} using a surprisingly
 high number of $5d$ holes $n_{\text{h}}>8$ are shown.}
 \label{Fig:msTc}
\end{figure}

In Fig.~\ref{Fig:msTc} we summarize currently published data for
the spin moment of $B'=$\,W,~Mo,~Re to investigate the relation of
Curie temperature $T_\textrm{C}$ and spin magnetic moment
$m_\textrm{S}$ at the $B'$ site of the ferrimagnetic double
perovskites of the type $A_2BB'$O$_6$. The data suggests that
indeed $T_\textrm{C}$ scales with $m_\textrm{S}$ at the $B'$ site.
Of course, we cannot draw final conclusions from this comparison
as other factors as structure, hybridization, and site occupation
can strongly influence $T_\textrm{C}$ \cite{Kato:04,DeTeresa:04}.
However, there is a clear trend that a large $T_\textrm{C}$ is
accompanied by a large magnetic moment $m_\textrm{S}$ at the $B'$
site. This finding is in qualitative agreement with the simple
model of ferrimagnetism mediated by itinerant minority spin
carriers. In the inset of Fig.~\ref{Fig:msTc} we show recent XMCD
data on other Re-based double perovskites which show the similar
trend \cite{Sikora:pre}. The relatively high values of the spin
magnetic moment for these samples are mainly due to the rather
high number of $d$-holes used in their data analysis. Nevertheless
our data and those of \cite{Sikora:pre} follow satisfactorily the
scaling law of Curie temperature and spin magnetic moment at the
$B'$ site. We emphasize that $T_{\rm C}$ seems not to depend on
the total spin moment given by the sum of the local moment on the
magnetic $B$ site (Cr, Fe) and the induced magnetic moment on the
non-magnetic $B^\prime$ site (Mo, W, Re). If this would be the
case, different $T_{\rm C}$ values would be expected for the Cr
and Fe based double perovskites in contrast to the experimental
data.

In summary, we have unambiguously demonstrated a large spin
magnetic moment at the Re $5d$ site in the ferrimagnetic double
perovskite Sr$_2$CrReO$_6$ together with a considerable orbital
magnetic moment. This result together with data from literature
suggest that the critical temperature for such double perovskites
scales with the magnetic moment at the 'non-magnetic' site of the
compound. The data is in fair agreement with the model prediction
and is important for understanding the mechanism of the high Curie
temperatures in the ferrimagnetic double perovskites.

This work was supported by the DFG (GR 1132/13), by the BMBF
(project 13N8279) and by the ESRF (HE-1882).

\end{document}